\documentclass[%
 reprint,
superscriptaddress,
 amsmath,amssymb,
 aps,
prb,
]{revtex4-2}
\usepackage[colorlinks=true, linkcolor=blue, citecolor=blue, urlcolor=blue]{hyperref}
\usepackage{graphicx}
\usepackage{hyperref}
\usepackage{dcolumn}
\usepackage{bm}
\usepackage[section]{placeins}

\begin{document}

\preprint{APS/123-QED}

\title{Three-dimensional real-space electron dynamics in graphene driven by strong laser fields}




\author{Siyuan Li}
\email{li-siyuan@g.ecc.u-tokyo.ac.jp}
\affiliation{%
 Department of Nuclear Engineering and Management, Graduate School of Engineering, The University of Tokyo,7-3-1 Hongo, Bunkyo-ku, Tokyo 113-8656, Japan
}

\author{Mizuki Tani}
\affiliation{%
 Kansai Institute for Photon Science, National Institutes for Quantum Science and Technology (QST), 8-1-7 Umemidai, Kizugawa, Kyoto 619-0215, Japan
}

\author{Arqum Hashmi}
\affiliation{%
 Department of Nuclear Engineering and Management, Graduate School of Engineering, The University of Tokyo,7-3-1 Hongo, Bunkyo-ku, Tokyo 113-8656, Japan
}

\author{Kenichi L. Ishikawa\thanks{Corresponding author: \email{ishiken@n.t.u-tokyo.ac.jp}}}
\email{ishiken@n.t.u-tokyo.ac.jp}
\affiliation{%
 Department of Nuclear Engineering and Management, Graduate School of Engineering, The University of Tokyo,7-3-1 Hongo, Bunkyo-ku, Tokyo 113-8656, Japan
}
\affiliation{%
 Photon Science Center, Graduate School of Engineering, The University of Tokyo, 7-3-1 Hongo, Bunkyo-ku, Tokyo 113-8656, Japan
}
\affiliation{%
 Research Institute for Photon Science and Laser Technology, The University of Tokyo, 7-3-1 Hongo, Bunkyo-ku, Tokyo 113-0033, Japan
}
\affiliation{%
Institute for Attosecond Laser Facility, The University of Tokyo, 7-3-1 Hongo, Bunkyo-ku, Tokyo 113-0033, Japan
}






\date{\today}

\begin{abstract}
We theoretically investigate the three-dimensional (3D) electron dynamics of graphene in real space under strong laser fields using time-dependent density functional theory (TDDFT). We successfully reproduce the reversal of current direction originating from the cancellation of two oppositely directed residual currents, as previously predicted by Morimoto \textit{et al.} [Y. Morimoto~\textit{et al.}, New J. Phys.~\textbf{24}, 033051 (2022)]. By distinguishing contributions from individual orbitals, our results validate the two-level system approximation and also emphasize that the first-principles approach agrees better with experimental results for light-driven residual current, especially in extremely strong fields. Furthermore, our 3D model reveals that the real-space atomic-scale current induced by strong laser fields is concentrated slightly above and below the graphene basal plane, rather than strictly within it. The two oppositely directed currents exhibit a pronounced height separation in the out-of-plane direction, indicating that the ring current is not confined to the graphene plane but forms a rotating 3D circulation loop which is absent in the reduced-dimensional model.
\end{abstract}

\maketitle


\section{\label{sec:level1}Introduction}

Since its discovery, graphene~\cite{novoselov2004electric} has attracted extensive attention due to the zero band gap, the weak screening effect, high laser damage threshold, and ultrafast optical response~\cite{geim2009graphene,castro2009electronic}.
These unique features make graphene one of the ideal materials for studying lightwave electronics, i.e., the ultrafast generation and manipulation of macroscopic electric currents through the waveform of ultrashort laser pulses.

In most studies, the electron dynamics in graphene (and solids in general) driven by strong laser fields is analyzed in the momentum (reciprocal) space~\cite{ishikawa2010nonlinear,ishikawa2013electronic, kelardeh2015graphene, taucer2017nonperturbative, heide2021optical}. 
The conduction and valence bands of graphene meet at the high-symmetry points K and $\mathrm{K}^{\prime}$, forming a characteristic Dirac cone.
The interband transitions preferentially occur around these points, and the intraband electron dynamics follows the acceleration theorem, $\bm{k}(t) = \bm{k_0} + e\bm{A}(t)/\hbar$, where $\bm{k_0}$ is the initial, field-free wavevector, $e$ the elementary charge, $\bm{A}(t)$ the vector potential of the laser field, and $\hbar$ the reduced Planck constant.

The nonlinear interaction of laser pulses with low temporal symmetry can induce controllable currents in dielectrics~\cite{franco2007robust,schiffrin2013optical,chen2018stark}. Similar field-driven current phenomena have also been explored in gapless materials such as graphene. Ishikawa has theoretically predicted the generation of a residual current in graphene under intense monocycle pulses~\cite{ishikawa2013electronic}. Then, Higuchi $\textit{et al.}$ have experimentally demonstrated that the residual electric current is generated by a spatially asymmetric few-cycle laser pulse because of the asymmetric carrier population in the reciprocal space induced via Landau-Zener-Stückelberg interference~\cite{higuchi2017light}.
They have also shown that its direction can be reversed by increasing the laser field \cite{higuchi2017light,heide2018coherent,weitz2024strong}. 

On the other hand, real-space electron dynamics is emerging as a powerful approach for intuitively understanding ultrafast and nonlinear phenomena under strong light fields~\cite{morimoto2022atomic,heide2024petahertz,yuan2024attosecond,morimoto2018diffraction,morimoto2024field}. More recently, Morimoto $\textit{et al.}$~\cite{morimoto2022atomic}, including one of the present authors, have theoretically studied the atomic-scale real-space distribution of the light-field-induced currents, by combining the nearest-neighbor tight-binding (TB) model and the semiconductor Bloch equation (SBE). 
It has been revealed that the atomic-scale currents flow dominantly through the network of the $\pi$ bonds. However, surprisingly, when the residual current changes its sign with increasing laser intensity, this flow does not vanish uniformly, but the back current appears through the center of the hexagonal atomic framework, canceling the current along the hexagon.
While the inclusion of the overlap integral between neighboring atomic orbitals, often neglected in the momentum-space analysis, was essential to properly describe the charge density between atomic sites, the same feature of the theoretical modeling led to slight violation ($\sim 10\%$) of the continuity equation for the current~\cite{morimoto2022atomic}.
This concern motivates us to perform a verification using first-principles calculations.


In the present study, we extend this previous work~\cite{morimoto2022atomic}, expected to contain the essential physics in the simplest possible manner, to verify its findings with the time-dependent density functional theory (TDDFT).
TDDFT is an \textit{ab initio} framework successfully applied to ultrafast electron dynamics under intense laser fields in molecules~\cite{casida1995time,castro2004excited, yabana2006real, adamo2013calculations, runge1984density,casida1995time,noda2019salmon} and solids~\cite{yabana1996time, nobusada2004high, otobe2008first, yabana2012time, kaneshima2018polarization, uemoto2019nonlinear, yamada2019energy, tani2021semiclassical, tani2022enhanced, hashmi2022nonlinear, hashmi2022valley, hashmi2023enhancement, otobe2024influence}. 
We systematically investigate graphene’s the real-space electron dynamics in three dimensions, which provides deeper insights into the spatial distribution of electron currents under intense laser fields.
Our numerical results well reproduce the measured dependence of the residual current on field strength and confirm the key results of Ref.~\cite{morimoto2022atomic}.
Moreover, our calculation reveals the current distribution perpendicular to the graphene plane. In particular, the ring current reported in the previous work ~\cite{morimoto2022atomic} is neither confined to the graphene plane nor even parallel to it but instead forms a rotating 3D circulation loop with pronounced out-of-plane height difference.


This paper is organized as follows. Section~\ref{TDDFT} briefly reviews TDDFT and how to evaluate the current. 
Section \ref{GSLR} shows the calculated ground state and linear response properties of graphene. 
In Sec.~\ref{Discussion} we present and discuss our numerical results. 
Conclusions are given in Sec.~\ref{conclusions}.

\section{Time dependent density functional theory}
\label{TDDFT}

The quantum dynamics of a many-electron system subject to a laser field is calculated by solving
the time-dependent Kohn-Sham equations (TDKS) for the time-dependent Bloch orbitals $u_{n,\mathbf{k}}(\mathbf{r}, t)$, with band index $n$ and crystalline momentum $\mathbf{k}$~\cite{bertsch2000real, otobe2008first}, which inherit the lattice periodicity of the system such that $u_{n,\mathbf{k}}(\mathbf{r} + \mathbf{R}, t) = u_{n,\mathbf{k}}(\mathbf{r}, t)$, where $\mathbf{R}$ is the lattice vector, TDKS reads,
%
\begin{widetext}
\begin{equation}
i \hbar \frac{\partial}{\partial t} u_{n,\mathbf{k}}(\mathbf{r}, t)= \left\{\frac{1}{2 m}\left(-i \hbar \nabla+\hbar \mathbf{k}+\frac{e}{c} \mathbf{A}(t)\right)^{2}-e\varphi(\mathbf{r},t) + \hat{v}_{\mathrm{NL}}^{\mathbf{k}+\frac{e}{\hbar c}\mathbf{A}(t)} + V_{\mathrm{xc}}\left[n(\mathbf{r},t)\right]\right\}u_{n,\mathbf{k}}(\mathbf{r}, t),
    \label{con:TDKS equation}
\end{equation}
\end{widetext}
where $m$ denotes the electron mass, $c$ the vacuum velocity of light, and $V_{\mathrm{xc}}\left[n(\textbf{r},t)\right]$ the exchange–correlation potential. The scalar potential $\varphi(\mathbf{r},t)$ includes the Hartree potential and the local part of norm-conserving ionic pseudopential~\cite{kleinman1982efficacious, fuchs1999ab}. The nonlocal part of the pseudopotential is defined as $\hat{v}_{\mathrm{NL}}^{\mathbf{k}} = e^{-i\mathbf{k}\cdot\mathbf{r}} \hat{v}_{\mathrm{NL}} e^{i\mathbf{k}\cdot\mathbf{r}}$, where $\hat{v}_{\mathrm{NL}}$ can be separated from the norm-conserving pseudopotential.

The time-dependent electron density $n(\mathbf{r},t)$ is given by,
\begin{equation}
 n(\mathbf{r},t)=\frac{1}{N_k}\sum_{\mathbf{k},n \in \mathrm{occ}} \left|u_{n,\mathbf{k}}(\mathbf{r}, t)\right|^2,
    \label{con: charge density}
\end{equation}
where $N_k$ and $\mathrm{occ}$ are the number of $k$ points and the occupied orbitals, respectively.
We adopt the Perdew–Zunger self-interaction-corrected local density approximation (LDA) functional for $V_{\mathrm{xc}}$~\cite{perdew1981self}, which, for the case of graphene, yields accuracy comparable to the generalized gradient approximation ~\cite{lazzeri2008impact} and, at the same time, can reduce the computational cost.

We evaluate the time-dependent electron current density distribution as, 
\begin{equation}
\begin{split}
 \mathbf{J}(\mathbf{r}, t)=&-\frac{e}{mN_k}\text{Re} \sum_{\mathbf{k},n \in \mathrm{occ}}u_{n,\mathbf{k}}^*(\mathbf{r}, t)\\ &\times\left(-i \hbar \nabla+\hbar \mathbf{k}+\frac{e}{c} \mathbf{A}(t)\right) u_{n,\mathbf{k}}(\mathbf{r}, t),    
\end{split}
    \label{con: current density distribution}  
\end{equation}
and its volume average as,
\begin{equation}
 \mathbf{I}(t)=\frac{1}{\Omega} \int_{\Omega} \mathbf{J}(\mathbf{r}, t) \mathrm{d} \mathbf{r} \, +\mathbf{I}_{\mathrm{NL}},
    \label{con: current density}  
\end{equation}
where $\Omega$ denotes the volume of the unit cell, and $\mathbf{I}_{\mathrm{NL}}$ accounts for the contribution from the nonlocal pseudopotential in the unit cell which can be calculated as~\cite{yamada2025saturable}:
\begin{equation}
\begin{split}
 \mathbf{I}_{\mathrm{NL}}(t)=&-e\int_{\Omega} \frac{1}{N_k}\sum_{\mathbf{k},n \in \mathrm{occ}}u_{n,\mathbf{k}}^*(\mathbf{r}, t)\\ 
 &\times\frac{1}{i\hbar}\left[\mathrm{r}, \hat{v}_{\mathrm{NL}}^{\mathbf{k}+\frac{e}{\hbar c}\mathbf{A}(t)} \right]  u_{n,\mathbf{k}}(\mathbf{r}, t).
\end{split}
    \label{con: nonlocal current}  
\end{equation}
We define the time-dependent occupancy $\rho_{n,\mathbf{k}}(t)$ through the projection of the time-dependent Bloch orbitals $u_{n,\mathbf{k}}(t)$ onto the initial field-free orbitals $u^\mathrm{GS}_{n,\mathbf{k}}$ as,
\begin{equation}
\rho_{n,\mathbf{k}}(t)=\sum_{m \in \mathrm{occ}} \left|\int_{\Omega} \mathrm{d} \mathbf{r} \, u_{m, \mathbf{k}}^{*}(\mathbf{r}, t) u_{n, \mathbf{k}}^{\mathrm{GS}}(\mathbf{r})\right|^{2}.
    \label{con: occupancy}  
\end{equation}

Using the open-source package SALMON~\cite{noda2019salmon}, our simulation is conducted on an orthogonal unit cell of graphene containing four carbon atoms, discretized into $24\times14\times85$ real-space and $45\times75\times1$ reciprocal-space grids. A vacuum layer with a total thickness of 1.5 nm (0.75 nm each side) perpendicular to the graphene plane is introduced. This calculation domain 
corresponds to the orthogonal lattice parameters as $a=0.426\, \text{nm}$, $b=0.246\,\text{nm}$, $c=1.5\,\text{nm}$.
Fourier interpolation is applied to smooth the real-space data for visualization.
To compare our results with those in Ref.~\cite{morimoto2022atomic}, we project the electron density and current density onto the graphene plane ($xy$ plane) as $n_{2D\_ z}(x,y)=\int_{0}^{c} n(\mathbf{r})\, \mathrm{d}z$, and $\mathbf{J}_{2D\_z}(x,y,t)=\int_{0}^{c} \mathbf{J}(\mathbf{r}, t) \,\mathrm{d}z$. 
Similarly, to analyze the out-of-plane current distribution, we project $n(\mathbf{r},t)$ and $\mathbf{J}(\mathbf{r}, t)$ onto the $zx$ plane as $n_{2D\_y}(x,z)=\int_{0}^{b} n(\mathbf{r})\, \mathrm{d} y \, $ and $\mathbf{J}_{2D\_y}(x,z,t)=\int_{0}^{b} \mathbf{J}(\mathbf{r}, t)\, \mathrm{d}y$.

\section{GROUND STATE AND LINEAR RESPONSE}
\label{GSLR}

\begin{figure}[tbp]
\includegraphics[width=8.0cm]{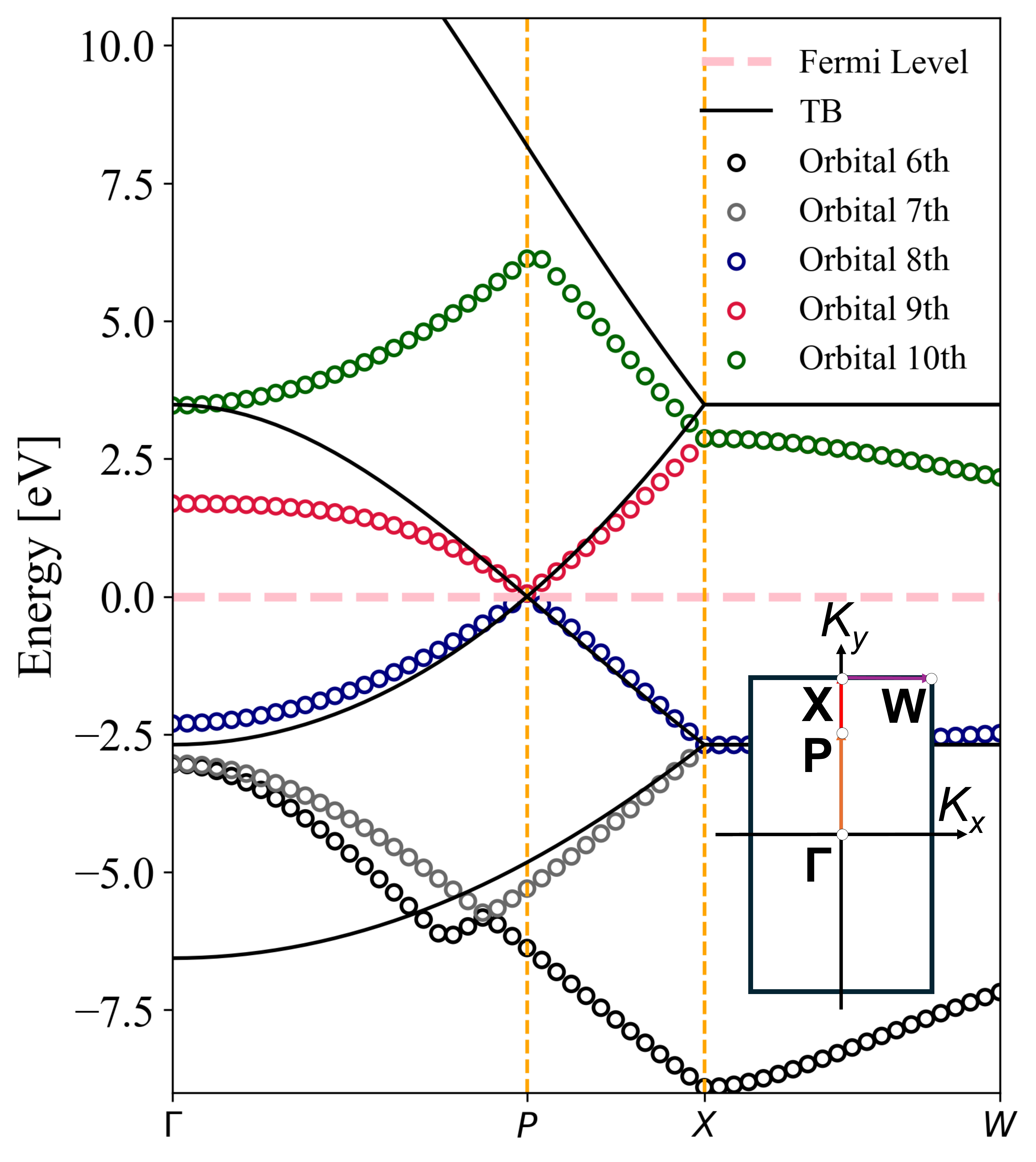} 
\caption{\label{fig:bsbg} Band structure of each orbital along high symmetry $\Gamma-P-X-W$ direction, for each orbital. Open circles represent the DFT results while the solid line indicates the tight-binding results.}
\end{figure}

The initial state of the system is taken as the ground state of the field-free Hamiltonian.
The calculated band structure is shown in Fig.~\ref{fig:bsbg}, which represents the energy eigenvalues of the Bloch orbital functions.
For comparison, the $\pi^*$ and $\pi$ bond energies $E_{\mathrm{c}}$ and $E_{\mathrm{v}}$, respectively,~\cite{suzuki2017electronic}
\begin{equation}
\begin{array}{c}
E_{\mathrm{c}}=\frac{t_{0}-t_{1} \sqrt{1+4 \cos ^{2}\left(\frac{a}{2} k_{y}\right) \pm 4 \cos \left(\frac{\sqrt{3} a}{2} k_{x}\right) \cos \left(\frac{a}{2} k_{y}\right)}}{1-s_{1} \sqrt{1+4 \cos ^{2}\left(\frac{a}{2} k_{y}\right) \pm 4 \cos \left(\frac{\sqrt{3} a}{2} k_{x}\right) \cos \left(\frac{a}{2} k_{y}\right)}}, \\
E_{\mathrm{v}}=\frac{t_{0}+t_{1} \sqrt{1+4 \cos ^{2}\left(\frac{a}{2} k_{y}\right) \pm 4 \cos \left(\frac{\sqrt{3} a}{2} k_{x}\right) \cos \left(\frac{a}{2} k_{y}\right)}}{1+s_{1} \sqrt{1+4 \cos ^{2}\left(\frac{a}{2} k_{y}\right) \pm 4 \cos \left(\frac{\sqrt{3} a}{2} k_{x}\right) \cos \left(\frac{a}{2} k_{y}\right)}},
\end{array}
\end{equation}
of the rectangular 4-atom unit cell tight-binding model are also plotted, with the on-site hopping parameter $t_0=0\, \mathrm{eV}$, the  adjacent hopping parameter $t_1=-3.033\, \mathrm{eV}$ and the overlap integral of $s_1=\,0.129$. 
We can clearly see the Dirac cone with zero band gap at the Fermi level and the $P$ point. 
The tight-binding model and DFT results closely agree with each other around there.
Since we use the orthogonal cell, the $\pi^*$ (9th orbital) local band separation at the $\Gamma$ point is smaller than for the case of the primitive cell due to band-folding effects.
The discrepancy between the DFT and TB $\pi^*$ energies at $\Gamma$ (TB: 6.17 eV, DFT: 3.99 eV) mainly originates from the absence of higher-order hopping terms in the TB model and the different treatment of the $\sigma$ band (7th orbital) contributions. In addition, near $\Gamma$, the $\sigma^*$ band (10th orbital) forms the shallow conduction-band minimum, whereas the $\pi^*$ band lies at much higher energy. This implies that when electrons are excited into regions far away from the Dirac cone, the resulting carrier dynamics deviate from a two-level description and exhibit pronounced multiband character.

\begin{figure}[tbp]
\includegraphics[width=8.0cm]{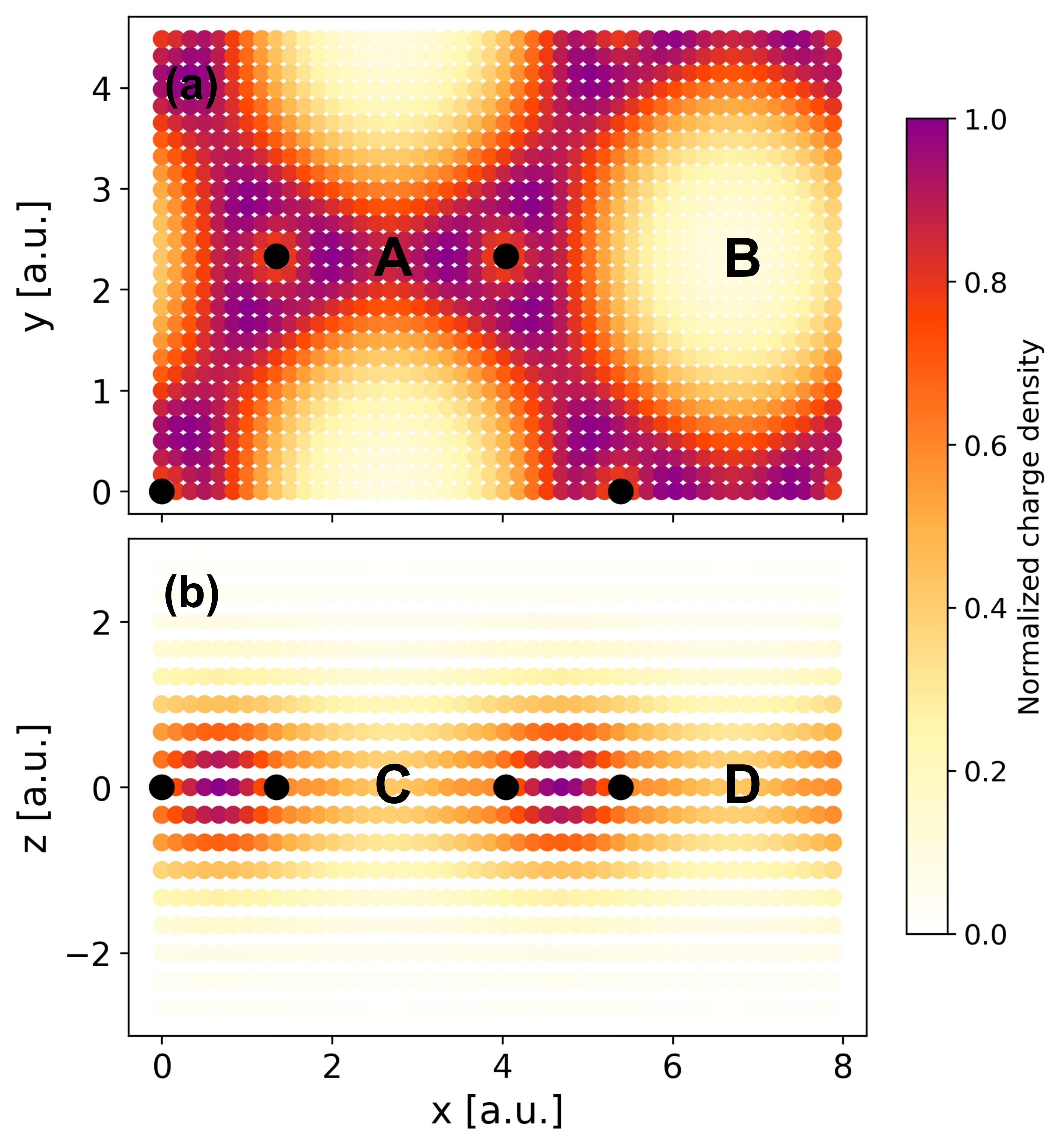} 
\caption{\label{fig:chargedensity} (a) Top view and (b) side view of normalized charge density in real space. A and B show position at the midpoint of a C–C bond and at the center of the hexagonal structure, respectively, projected onto the $xy$ plane. C and D correspond to the midpoint of a C–C bond oriented at 0$^\circ$ and 60$^\circ$ with respect to the x-axis, respectively. Black circles indicate carbon nuclei. All densities are normalized.}
\end{figure}

The charge density projected on $xy$ and $zx$ planes in real space is displayed in Fig.~\ref{fig:chargedensity}. Electrons are concentrated in the hexagonal frame of graphene while the hexagonal center has the lowest charge density [Fig.~\ref{fig:chargedensity} (a)]. In Fig.~\ref{fig:chargedensity} (b), we see that the electron density is negligibly small at $z\sim \pm 2 $ a.u., compared to that on the basal plane. This observation is consistent with the experimental results~\cite{xu2011giant}.

The linear response is obtained by time propagation after an impulsive momentum kick, realized by a steplike vector potential, 
\begin{equation}
\mathbf{A}(t)=\left\{\begin{array}{ll}
-\mathbf{A}_{0} & (t \geqslant 0), \\
0 & (t<0),
\end{array}\right.
    \label{con: linearvectorpotential}
\end{equation}
which corresponds to an impulsive electric field $\mathbf{E}(t)=\mathbf{A}_0\delta(t)$. 
Since such an electric field has a constant power spectrum over all frequencies,
the diagonal component of the optical conductivity $\sigma_d(\omega)$ is calculated as,
\begin{equation}
\sigma_d(\omega) = -\frac{e}{|\mathbf{A}_0|} I_d(\omega) \quad (d = x, y, z),
    \label{con: conductivity}
\end{equation}
where $I_d(\omega)$ denotes the component along the polarization direction $d$ of the temporal Fourier transform of the volume-averaged current density.
The dielectric function $\varepsilon_d(\omega)$ is given by,
\begin{equation}
\varepsilon_d(\omega)=1+4\pi i\frac{\sigma_d(\omega)}{\omega}.
    \label{con: dielectric function}
\end{equation}
The off-diagonal components of the optical response tensors vanish due to the crystal symmetry of graphene.

\begin{figure}[tbp]
\includegraphics[width=8.0cm]{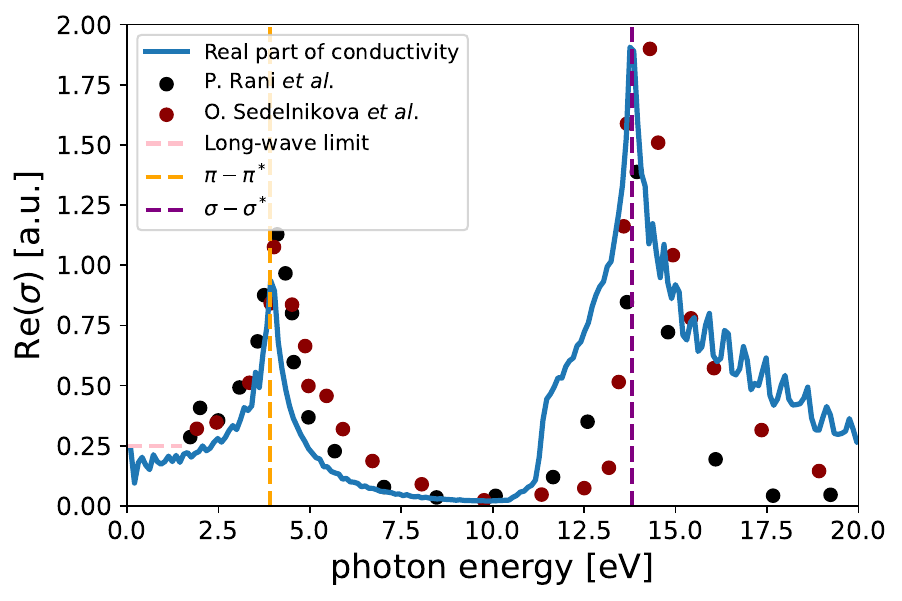} 
\caption{\label{fig:linear} Linear response of graphene computed with TDDFT. The green line shows the real part of conductivity. Black and red dots are results from DFT calculations~\cite{rani2014dft,sedelnikova2011ab}. The horizontal pink dashed line indicates the zero-frequency conductivity ($\sim$0.25 a.u.). The orange and purple vertical lines mark the two peak positions, corresponding to the $\pi\to\pi^*$ and $\sigma\to\sigma^*$ transitions, respectively.}
\end{figure}

Figure~\ref{fig:linear} shows the real-part of thus calculated conductivity, in comparison with previous DFT calculations~\cite{rani2014dft,sedelnikova2011ab}. 
Two distinct absorption peaks are present, corresponding to transitions between the $\pi$ and $\pi^*$ bands and between the $\sigma$ and $\sigma^*$ bands. At the $\Gamma$ point, the linear optical response is weak. In the long-wavelength limit ($\omega \to 0$), the conductivity approaches approximately 0.25 a.u., consistent with previous studies and experiments~\cite{kuzmenko2008universal, falkovsky2007space}. This agreement demonstrates the reliability of our TDDFT approach for linear optical properties.

\section{Results and Discussion}\label{Discussion}
We specifically consider a few-cycle CEP laser pulse with the following form:
\begin{equation}
E(t)=F\mathrm{sin}^4\left(\frac{\pi t}{\tau}\right)\mathrm{cos}(\omega t + \varphi_{CEP}),
    \label{con: waveform}
\end{equation}
where we vary the peak field amplitude $F$ up to 4 V/nm, within the reported damage threshold~\cite{roberts2011response}. 
The pulse foot-to-foot duration is $\tau=21$ fs, which corresponds to the full-width-at-half-maximum (FWHM) duration of 5.5 fs, the carrier-envelope phase $\varphi_{CEP}=\pi/2$, and $\omega = 2\pi\times0.375$ PHz (1.55 eV photon energy, 800 nm wavelength). A $\sin^4$ envelope is chosen to suppress high-frequency tails of the pulse~\cite{yamada2019energy}. The laser polarization is along a graphene C–C bond (taken as the +$x$ direction in our coordinates, see Fig.~\ref{fig:chargedensity}). The time step is set to be 0.02 a.u., and the total simulation time $T$ is 100 fs.

\begin{figure}[htbp]
\includegraphics[width=8cm]{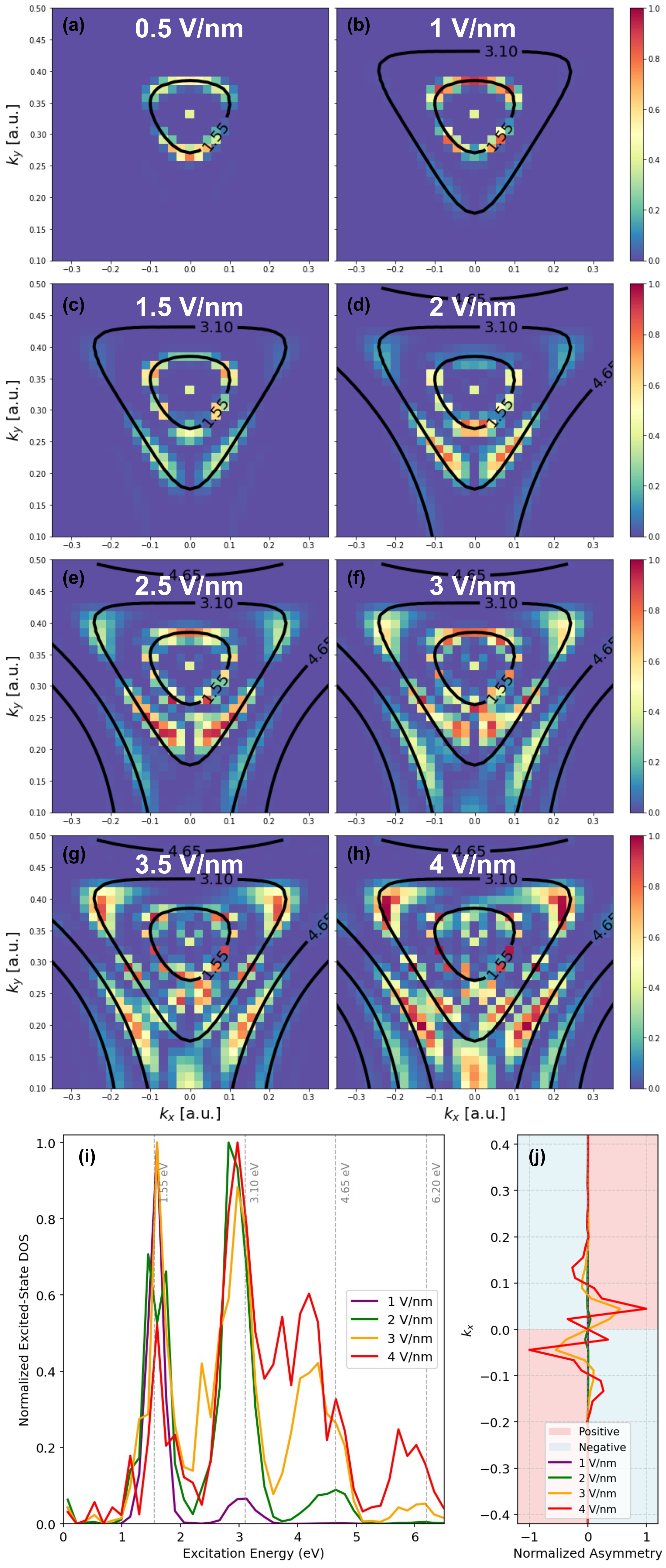} 
\caption{\label{fig:numberstate} The excited population $P(k_x,k_y)$ on the reciprocal space after the laser fields. (a-h) Population near Dirac cone on the lowest conduction orbital (9th orbital) with contour lines on times of photon energy under the peak field amplitudes from 0.5 V/nm to 4 V/nm. (i) Normalized density of states for the excited population on conduction band. (j) excited carrier distribution asymmetry $\mathcal{A}(k_{x_0})$ normalized to its maximum value on 4 V/nm as a function of $k_x$. Positive (red) and negative (blue) background shadings indicate the sign of the asymmetry.}
\end{figure}

Let us first focus on electron excitation in the reciprocal space after the laser pulse ($t = 21$ fs). Excited-state analysis is carried out by projecting the time-dependent Kohn-Sham orbitals onto Houston functions~\cite{houston1940acceleration}. Figure~\ref{fig:numberstate}(a)-(h) shows the calculated band occupation of varying laser field intensity. 
At low intensity, single-photon and then two-photon resonant excitation lead to carrier distribution symmetric with respect to the $k_y$ axis, i.e., $k_x=0$ [Fig.~\ref{fig:numberstate}(a)-(c)]. 
As the laser intensity increases, the region between the contour lines of single- and two-photon excitation starts to be populated [Fig.~\ref{fig:numberstate}(d)], indicative of tunneling excitation, and asymmetry in excited carrier distribution appears in that region [Fig.~\ref{fig:numberstate}(e)(f)], 
consistent with the observations of Ref.~\cite{higuchi2017light,morimoto2022atomic}. 

At even stronger fields ($3-4$ V/nm) [Fig.~\ref{fig:numberstate}(g)(h)], higher-energy regions between the multi-photon resonant excitation contours are filled [see also Fig.~\ref{fig:numberstate}(i)], which indicates increase in tunneling excitation. 
Simultaneously, the asymmetry extends to the higher-energy region. Figure~\ref{fig:numberstate}(j) shows the asymmetry in carrier distribution between $k_x > 0$ and $k_x < 0$ defined as $\mathcal{A}(k_{x}) = \sum_{k_y>0} \left[ P(k_{x}, k_y) - P(-k_{x}, k_y) \right]$. At 4 V/nm, the asymmetry $\mathcal{A}(k_{x})$ around $k_x=-0.02$ near the Dirac cone and $k_x=-0.13$ far from there which contribute to the current on negative direction surpasses the positive component around $k_x=0.04$, which leads to the negative total current.

In addition to the intraband residual current $I^{\text{intra-rd}}_{i}$ on band $i$, 
\begin{equation}
I^{\text{intra-rd}}_i = 
\left\{
\begin{aligned}
&\int_{\text{BZ}} \frac{\partial E_{i}}{\partial \mathbf{k}} \, \rho_{i} \, \mathrm{d}\mathbf{k}, & i \in {\rm conduction}, \\
&\int_{\text{BZ}} \frac{\partial E_{i}}{\partial \mathbf{k}} \, (2 - \rho_{i}) \, \mathrm{d}\mathbf{k}, & i \in {\rm valence,}
\end{aligned}
\right.
\label{con: intraband}
\end{equation}
we also evaluate the residual current by averaging the current after the pulse, which directly corresponds to the experimental measurement, as,
\begin{equation}
I_{d}^{\text{rd}}=\frac{1}{T-\tau}\int_{\tau}^{T}I_d(t)\,\mathrm{d}t,
    \label{con: rsdcurrent}
\end{equation}
where, $E_i$ and $\rho_i$ denote the band energy and the electron occupation after the pulse [Eq.~(\ref{con: occupancy})], respectively, of band $i$.
$I_{d}^{\mathrm{rd}}$ the average residual current density in $d$ direction and $I_d$ the component on $d$ direction of the volume average of current density $\mathbf{I}$. Similarly, the residual current distribution projected on $xy$ and $zx$ plane can be expressed as $\mathbf{J}_{2D\_z}^{\text{rd}}(x,y)=\frac{1}{T-\tau}\int_{\tau}^{T}\mathbf{J}_{2D\_z}(x,y,t)\,\mathrm{d} t$ and $\mathbf{J}_{2D\_y}^{\text{rd}}(x,z)=\frac{1}{T-\tau}\int_{\tau}^{T}\mathbf{J}_{2D\_y}(x,z,t)\,\mathrm{d}t$, respectively.

\begin{figure}[!tbp]
\includegraphics[width=8cm]{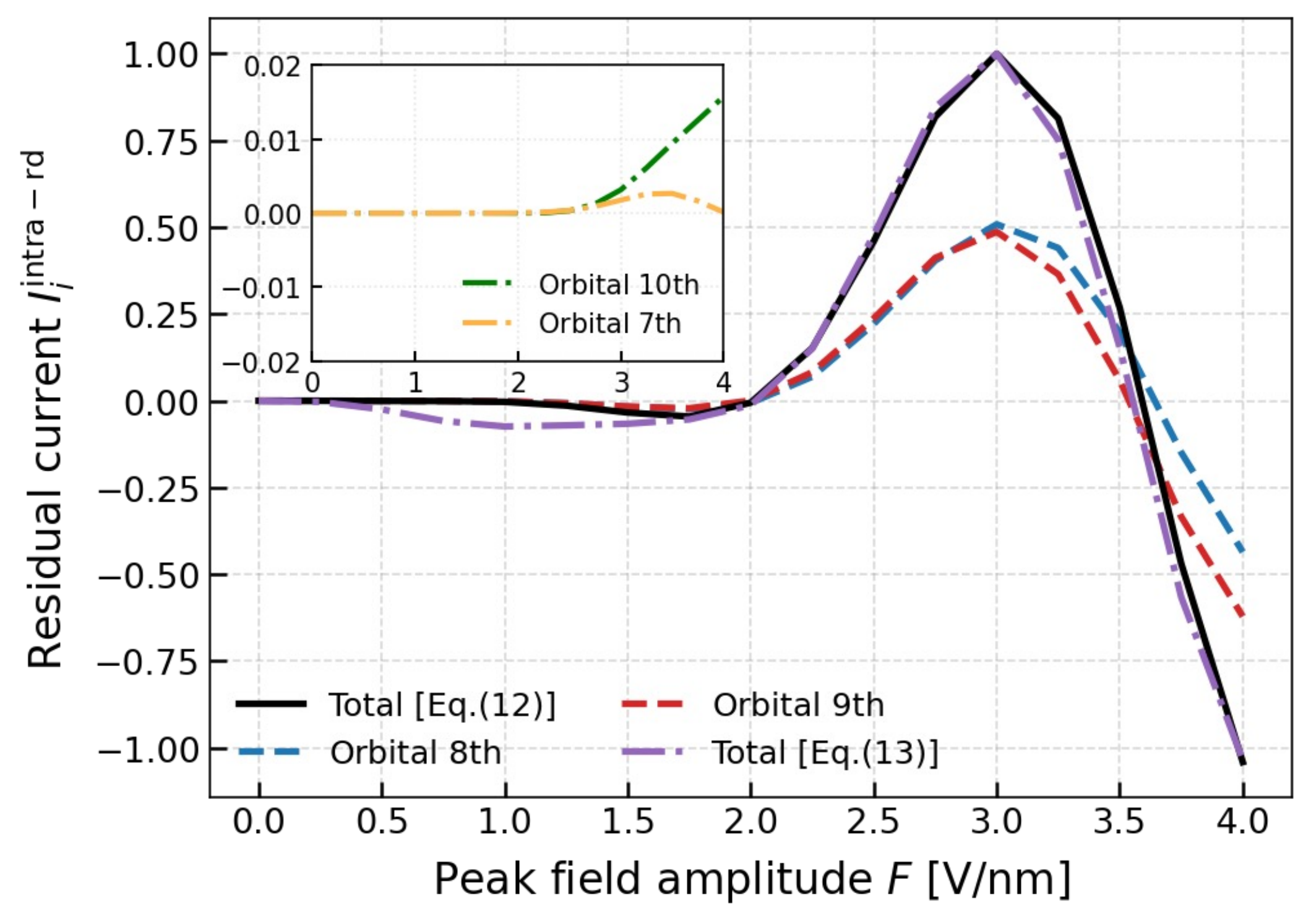}
\caption{\label{fig:vrsd} Normalized intraband residual current contribution from each orbital of graphene. Black solid line: total current calculated by summing the currents evaluated as Eq.~(\ref{con: intraband}) over all the bands, purple dashed-dotted line: the total current calculated via Eq.~(\ref{con: rsdcurrent}). The contributions from the 8th (valence top, blue line) and 9th (conduction bottom, red line) orbitals are also plotted. Inset: contributions from the 7th (orange line) and 10th (green line)  orbitals.}
\end{figure}

We have confirmed that our simulation time (100 fs) is sufficiently long that the contribution from the oscillating interband polarization is removed after the temporal integration in Eq.~(\ref{con: rsdcurrent}). As a result, the time-average residual current [Eq. (\ref{con: rsdcurrent})] becomes practically the same as the sum of the intraband residual currents [Eq. (\ref{con: intraband})] over all the orbitals, i.e., the remaining direct current (Fig.~\ref{fig:vrsd}).

\begin{figure}[!tbp]
\includegraphics[width=8.0cm]{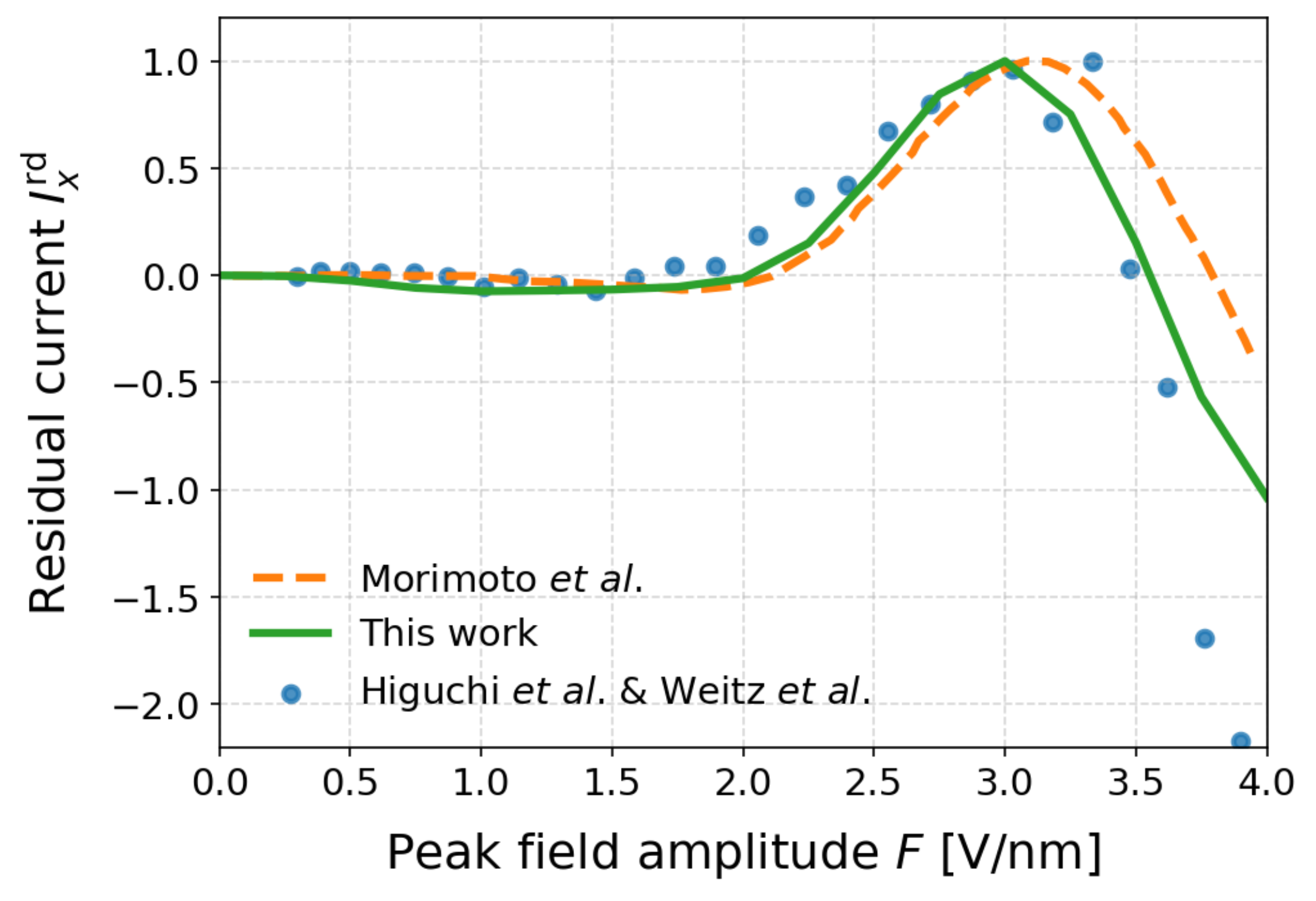} 
\caption{\label{fig:totalrsd} Total residual current vs.~peak field amplitude $F$. Green solid line: our TDDFT results (this work), orange dashed: calculated results by Morimoto $\textit{et al.}$~\cite{morimoto2022atomic}, blue circles: experimental data from Higuchi $\textit{et al.}$~\cite{higuchi2017light}  ($\leq$ 3 V/nm) extended by Weitz $\textit{et al.}$~\cite{weitz2024strong}. The plots are normalized by the maximum of the residual current, respectively.}
\end{figure}

In Fig.~\ref{fig:totalrsd} we show the residual current calculated with Eq.~(\ref{con: rsdcurrent}) as a function of field intensity.
Our TDDFT results agree very well with those calculated by the two-level SBE with the tight-binding model  \cite{morimoto2022atomic} and the experimental data \cite{higuchi2017light,weitz2024strong}.
In the strong-field regime ($F\geq 3\,\mathrm{V/nm}$), especially, our first-principles results lie closer to the experimental values than the SBE results do, which indicates that TDDFT describes nonlinear dynamics of highly excited electrons more accurately through the time-dependent effective Hamiltonian.
The direction of the residual current is reversed at $F\approx 2$ and $3.56 \,\mathrm{V/nm}$,
and the residual current reaches its maximum around 3 V/nm.

Figure~\ref{fig:vrsd} breaks down the intraband residual current by orbitals evaluated as Eq.~(\ref{con: intraband}). We find that the lowest conduction band (9th orbital) and the highest valence band (8th orbital) dominate the residual current. The contributions from these two orbitals are nearly equal (symmetric) up to $\approx 3$ V/nm field strength, indicating that excitation is confined within the Dirac cone. On the other hand, once the peak field exceeds 3 V/nm, part of electrons are driven into the energy regions beyond the pure Dirac cone where the 8th orbital and the 9th orbital are not symmetric [also see Fig.~\ref{fig:numberstate} (f)-(h)], leading to the difference between the valence-band and conduction-band currents. At the same time, occupation of higher bands (for instance, 10th orbital) becomes non-negligible, indicating that driven by strong fields, bands beyond the simple two-level model begin to contribute to the current.

\begin{figure}[!tbp]
\includegraphics[width=8.0cm]{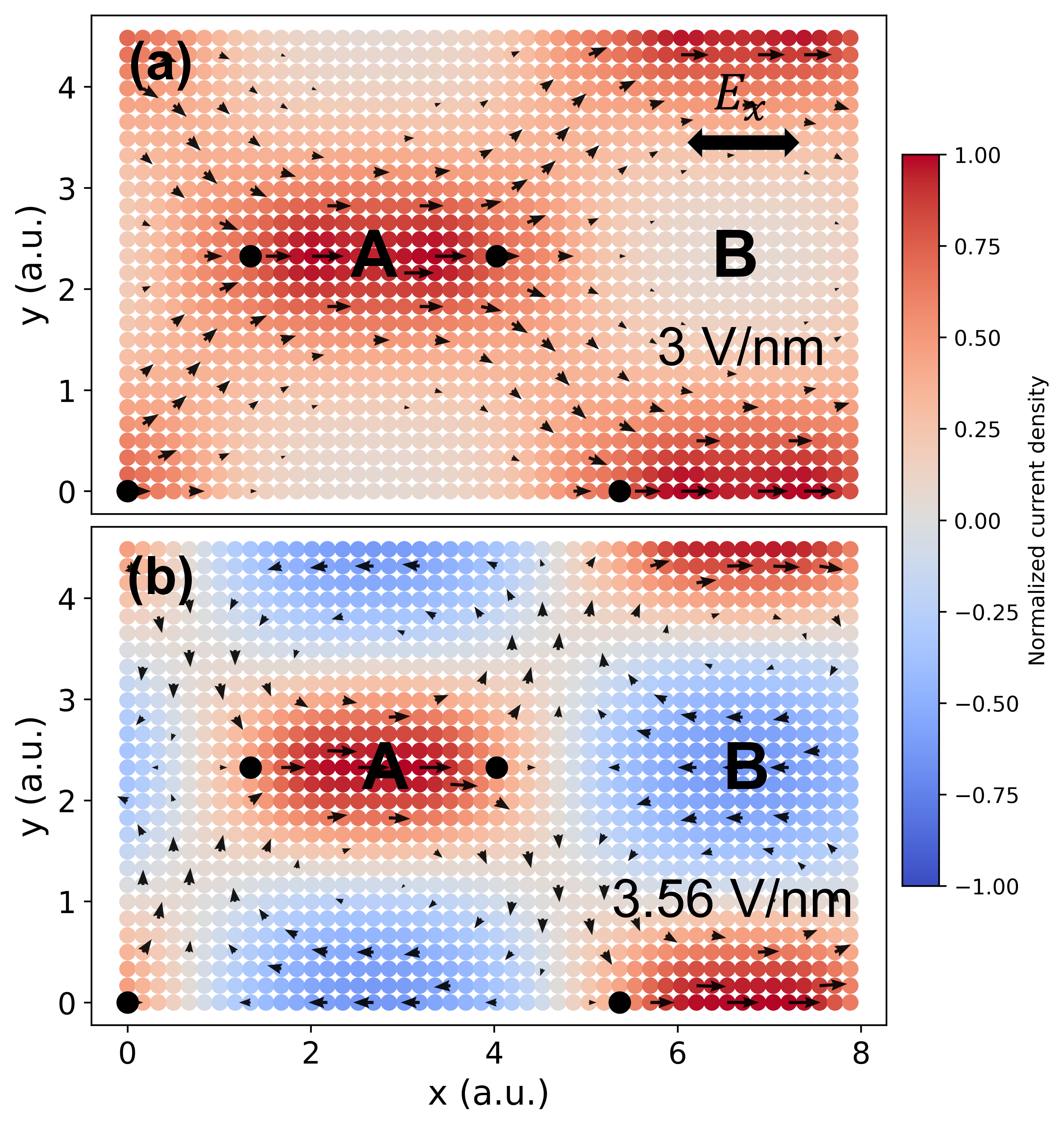} 
\caption{\label{fig:rsdsq} Normalized real-space distribution of the $x$ component of the local residual current projected onto the $xy$ plane. (a) peak field amplitudes of 3 V/nm (b) peak field amplitudes of 3.56 V/nm. Black arrows indicate the projected local current vectors, and black circles mark the carbon nuclei positions. (The nonlocal pseudopotential contribution to the current is not included in these plots.)}
\end{figure}

In Fig.~\ref{fig:rsdsq} we visualize the residual current distribution after the pulse in the real space, projected onto the graphene plane. 
At $F=3$ V/nm, at which the total residual current reaches its maximum  (see Fig.~\ref{fig:totalrsd}), the current flows along the hexagonal atomic framework, i.e., along the C–C bonds, especially along the laser polarization [Fig.~\ref{fig:rsdsq}(a)]. At $F=3.56$ V/nm, where the total residual current is nearly zero, the microscopic local current does not vanish, but a return current appears at the center of each hexagon, 
forming two ring currents within a single hexagon. 
These observations are consistent with our previous study~\cite{morimoto2022atomic}.

\begin{figure}[!tbp]
\includegraphics[width=8cm]{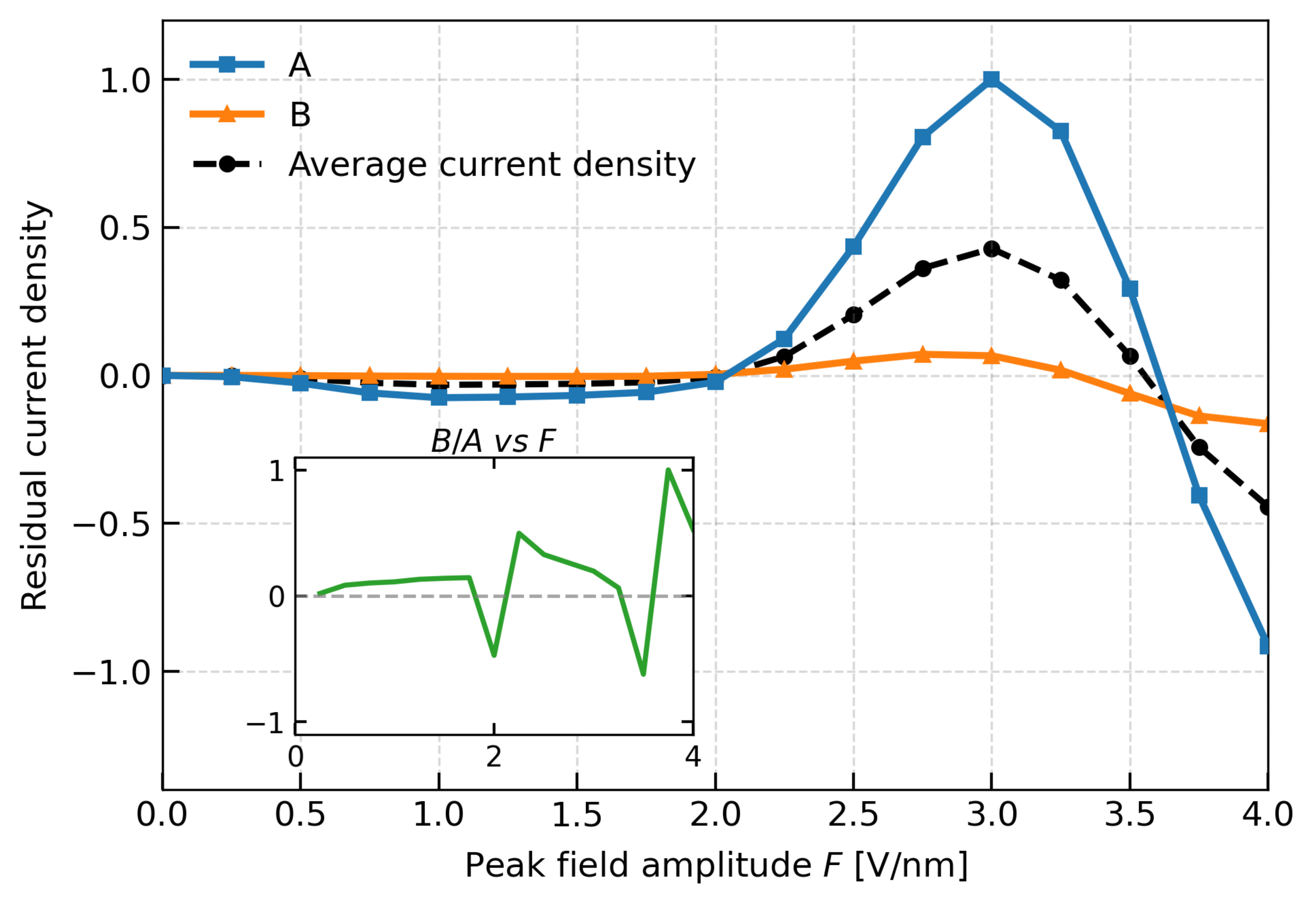}
\caption{\label{fig:ABrsd} Residual current density at two representative points in real space as a function of peak field amplitude: point A (midpoint of a C–C bond, along polarization) and point B (center of a hexagon). The blue curve is the current density at point A and the orange curve at point B. The black dashed line indicates the total residual current density (from Fig.~\ref{fig:totalrsd}) for reference. Inset: Normalized ratio of the residual current density at B and A as a function of the peak field amplitude $F$. The ratio takes negative values around $F \approx 2$ V/nm and $F \approx 3.56 $ V/nm.}
\end{figure}

Let us further examine the residual current density at two representative positions: point A [atomic unit coordinate (2.68, 2.32)] at the midpoint of a C–C bond (aligned with the field polarization) and point B [atomic unit coordinate (6.71, 2.32)] at the center of a hexagon (see Fig.~\ref{fig:rsdsq}). Figure~\ref{fig:ABrsd} plots the residual current density at these points as a function of peak field amplitude. Overall, the current densities at points A and B follow the same trend as the total residual current. If we look more closely, however, the local current at point B reverses its direction at a lower field ($\sim3.3$ V/nm) than the current at point A ($\sim3.6$ V/nm). As a consequence, the currents at the two points are in opposite directions at $F=3.56$ V/nm, which leads to the ring currents we have observed above in Fig.~\ref{fig:rsdsq}(b).

\begin{figure}[!tbp]
\includegraphics[width=8cm]{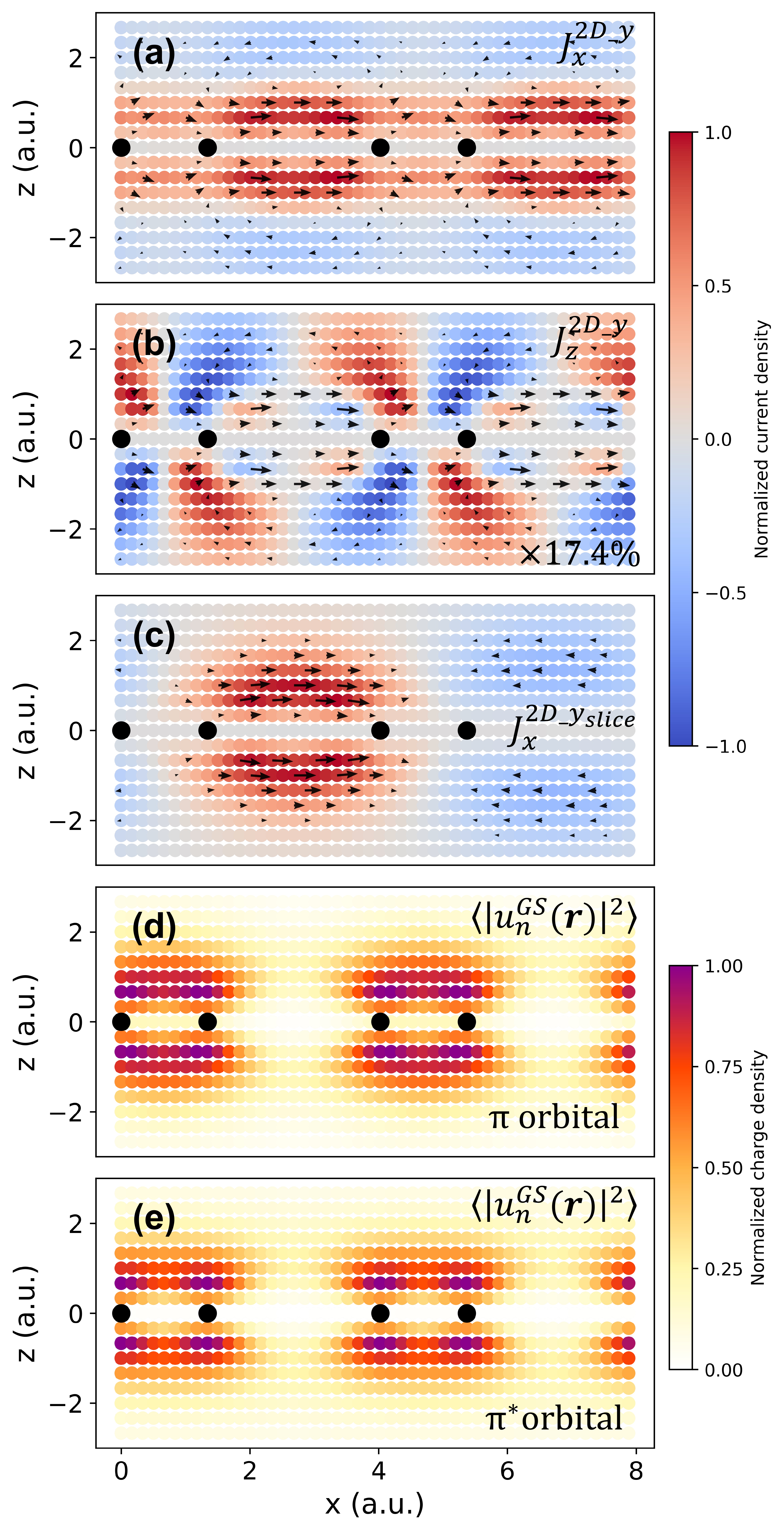}
\caption{\label{fig:3Dzrsd} Real-space distribution of the local residual current and ground-state orbital electron projected $zx$ plane. (a) Normalized real-space distribution of the $x$ component of the local residual current projected onto the $zx$ plane for a peak field amplitude of 3.56 V/nm. (b) Normalized real-space distribution of the $z$ component of the local residual current projected onto the $zx$ plane under the same field. $\times$17.4\% shows the color scaling factors in comparable with the maximum of (a).  (c) Normalized cross-sectional slice at $y_0 = 2.32$ a.u. of (a) (aligned with the A and B point in Fig.~\ref{fig:rsdsq}(b)). (d,e) The normalized ground-state charge densities of $\pi$ and $\pi^*$ Bloch orbitals projected onto the $zx$ plane. As in Fig.~\ref{fig:rsdsq}, black arrows indicate the projected local current vectors, and black circles mark the carbon nuclei positions.}
\end{figure}

So far, we have discussed the current distribution projected onto the graphene plane. We now turn to the three-dimensional current distribution. Figure~\ref{fig:3Dzrsd} presents the current density distribution projected onto the $zx$ plane for $F=3.56$ V/nm, which corresponds to Fig.~\ref{fig:rsdsq}(b). 
Out-of-plane circulating current is observed, with a positive current near the basal plane and an oppositely directed current emerging slightly away from the plane in Fig.~\ref{fig:3Dzrsd}(a,b). 
To clearly elucidate the 3D location of the return currents appearing in Fig.~\ref{fig:rsdsq}(b), we present in Fig.~\ref{fig:3Dzrsd}(c) $zx$-plane slice of the three-dimensional residual current density evaluated at $y_0 = 2.32$ a.u. where is the line connecting A and B in Fig.~\ref{fig:rsdsq}(b) as $\mathbf{J}^{\text{rd}}zx(x,z;y_0) = \frac{1}{T-\tau}\int_{\tau}^{T}\mathbf{J}(x,y_0,z, t) \,\mathrm{d}t$. The opposite current components are visible, both slightly shifted away from the graphene basal plane but with a subtle height difference at $z$ axis. This trend is consistent with the distribution in Fig.~\ref{fig:3Dzrsd}(b), and corroborates that the current at point B in Fig.~\ref{fig:rsdsq} resides at a comparatively more out-of-plane position.
Notably, virtually no current flows exactly in the graphene plane ($z=0$). Instead, the current is mainly concentrated approximately 1 a.u. above and below it, which 
coincides very well with the positions of the ground-state $\pi$ and $\pi^*$ orbitals [Fig.~\ref{fig:3Dzrsd}(d,e)]. 
This supports that 
the electron dynamics is driven predominantly in the $\pi$ and $\pi^*$ orbitals, as was assumed in the previous studies~\cite{morimoto2022atomic}. 
From the 3D view, the ring current reported in the previous work \cite{morimoto2022atomic} is neither confined to the graphene plane nor parallel to it but forms a rotating 3D circulation loop with pronounced out-of-plane height difference.


\section{Conclusions}
\label{conclusions}

We have performed first-principles real-time TDDFT simulations of graphene subject to intense few-cycle laser pulses. The calculated residual currents excellently reproduce the experimentally observed field-strength dependence, including the reversal of current direction. By resolving the contributions from individual orbitals, we have verified that the two-level description \cite{higuchi2017light,morimoto2022atomic} around the Dirac cone provides an adequate picture at moderate fields, while additional bands start to be relevant in the strong-field regime.

Our \textit{ab initio} results confirm and extend the scenario proposed in our previous work \cite{morimoto2022atomic}, in which current reversal arises from the cancellation of microscopic flows along the C-C bonds and at the center of the hexagon. The present TDDFT analysis not only validates this mechanism from \textit{ab initio} perspective, but also reveals that the dominant residual current is carried by $\pi$ and $\pi^*$ orbitals slightly above and below the basal plane, rather than strictly within it. Further, rather than being confined or parallel to the graphene plane, the out-of-plane projection reveals that this vortex-like current forms a rotating three-dimensional circulation loop. These findings connect the reciprocal- and three-dimensional real-space viewpoints of strong-field dynamics in graphene. 
Our work thus provides microscopic insight useful for the development of ultrafast optoelectronics and light-wave electronics with a more realistic interpretation beyond 2D descriptions.

\begin{acknowledgments}
This research was supported by MEXT Quantum Leap Flagship Program (MEXT Q-LEAP) under Grant No. JPMXS0118067246, JSPS KAKENHI Grants No. JP20H05670, No. JP22K18982, No. JP24H00427, No. JP24K01224, No. JP24K23024, No. JP25K17367, No. 25H00704 and JST SPRING, Grant Number JPMJSP2108. This work was partially supported by the RIKEN TRIP initiative. S. Li acknowledges support from WINGS-QSTEP Program and the Special Graduate Program in Resilience Engineering (Nuclear Engineering and Management Course), the University of Tokyo. This research was also supported by SEUT Grant for International Students under the University of Tokyo Graduate School of Engineering. The numerical calculations are carried out using the computer facilities of Wisteria at the University of Tokyo (Project ID No. go73). We are grateful to Yuya Morimoto for the fruitful discussions.

The authors declare no competing financial interest.
\end{acknowledgments}






\newpage
\bibliography{apssamp}

@article{lazzeri2008impact,
  title={Impact of the electron-electron correlation on phonon dispersion: Failure of LDA and GGA DFT functionals in graphene and graphite},
  author={Lazzeri, Michele and Attaccalite, Claudio and Wirtz, Ludger and Mauri, Francesco},
  journal={Phys. Rev. B},
  volume={78},
  number={8},
  pages={081406},
  year={2008},
  publisher={APS},
  url = {https://journals.aps.org/prb/abstract/10.1103/PhysRevB.78.081406}
}

@article{higuchi2017light,
  title={Light-field-driven currents in graphene},
  author={Higuchi, Takuya and Heide, Christian and Ullmann, Konrad and Weber, Heiko B and Hommelhoff, Peter},
  journal={Nature},
  volume={550},
  number={7675},
  pages={224--228},
  year={2017},
  publisher={Nature Publishing Group UK London},
  url = {https://doi.org/10.1038/nature23900}
}

@article{morimoto2022atomic,
  title={Atomic real-space perspective of light-field-driven currents in graphene},
  author={Morimoto, Yuya and Shinohara, Yasushi and Ishikawa, Kenichi L and Hommelhoff, Peter},
  journal={New J. of Phys.},
  volume={24},
  number={3},
  pages={033051},
  year={2022},
  publisher={IOP Publishing},
  url ={https://iopscience.iop.org/article/10.1088/1367-2630/ac5c18/meta}
}

@article{noda2019salmon,
  title={SALMON: Scalable ab-initio light--matter simulator for optics and nanoscience},
  author={Noda, Masashi and Sato, Shunsuke A and Hirokawa, Yuta and Uemoto, Mitsuharu and Takeuchi, Takashi and Yamada, Shunsuke and Yamada, Atsushi and Shinohara, Yasushi and Yamaguchi, Maiku and Iida, Kenji and others},
  journal={Comput. Phys. Commun.},
  volume={235},
  pages={356--365},
  year={2019},
  publisher={Elsevier},
  url = {https://www.sciencedirect.com/science/article/pii/S0010465518303412?casa_token=8PMvGYTerdEAAAAA:Pb3BzlUOKZ6D6Pk9HMTvDnrV2Az8i7JqyNy-MGAZ2lEVHcL0tsk0L-D52i4accM1YaqlrRvy7Eo}
}

@article{roberts2011response,
  title={Response of graphene to femtosecond high-intensity laser irradiation},
  author={Roberts, Adam and Cormode, Daniel and Reynolds, Collin and Newhouse-Illige, Ty and LeRoy, Brian J and Sandhu, Arvinder S},
  journal={Appl. Phys. Lett.},
  volume={99},
  number={5},
  year={2011},
  publisher={AIP Publishing},
  url = {https://pubs.aip.org/aip/apl/article-abstract/99/5/051912/374257/Response-of-graphene-to-femtosecond-high-intensity?redirectedFrom=fulltext}
}

@article{rani2014dft,
  title={DFT study of optical properties of pure and doped graphene},
  author={Rani, Pooja and Dubey, Girija S and Jindal, VK},
  journal={Physica E},
  volume={62},
  pages={28--35},
  year={2014},
  publisher={Elsevier},
  url = {https://www.sciencedirect.com/science/article/pii/S1386947714001374?casa_token=BsY8rcsIdlQAAAAA:dHeipvnnkCorZThtKdLhOnKIVOvec8FQvPKuRrVQ92aB5y35J072OCNHE5vnsVF-bplhCOWgRf4}
}

@article{sedelnikova2011ab,
  title={Ab initio study of dielectric response of rippled graphene},
  author={Sedelnikova, OV and Bulusheva, LG and Okotrub, AV},
  journal={J. Chem. Phys.},
  volume={134},
  number={24},
  year={2011},
  publisher={AIP Publishing},
  url = {https://pubs.aip.org/aip/jcp/article-abstract/134/24/244707/1004648/Ab-initio-study-of-dielectric-response-of-rippled?redirectedFrom=fulltext}
}

@article{yabana2006real,
  title={Real-time, real-space implementation of the linear response time-dependent density-functional theory},
  author={Yabana, K and Nakatsukasa, T and Iwata, J-I and Bertsch, GF},
  journal={Phys. Status Solidi (B)},
  volume={243},
  number={5},
  pages={1121--1138},
  year={2006},
  publisher={Wiley Online Library},
  url = {https://onlinelibrary.wiley.com/doi/abs/10.1002/pssb.200642005}
}

@article{nobusada2004high,
  title={High-order harmonic generation from silver clusters: Laser-frequency dependence and the screening effect of d electrons},
  author={Nobusada, Katsuyuki and Yabana, Kazuhiro},
  journal={Phys. Rev. A},
  volume={70},
  number={4},
  pages={043411},
  year={2004},
  publisher={APS},
  url = {https://journals.aps.org/pra/abstract/10.1103/PhysRevA.70.043411}
}

@article{otobe2008first,
  title={First-principles electron dynamics simulation for optical breakdown of dielectrics under an intense laser field},
  author={Otobe, T and Yamagiwa, M and Iwata, J-I and Yabana, K and Nakatsukasa, T and Bertsch, GF},
  journal={Phys. Rev. B},
  volume={77},
  number={16},
  pages={165104},
  year={2008},
  publisher={APS},
  url = {https://journals.aps.org/prb/abstract/10.1103/PhysRevB.77.165104}
}

@article{uemoto2019nonlinear,
  title={Nonlinear polarization evolution using time-dependent density functional theory},
  author={Uemoto, Mitsuharu and Kuwabara, Yuki and Sato, Shunsuke A and Yabana, Kazuhiro},
  journal={J. Chem. Phys.},
  volume={150},
  number={9},
  year={2019},
  publisher={AIP Publishing},
  url = {https://pubs.aip.org/aip/jcp/article-abstract/150/9/094101/76133/Nonlinear-polarization-evolution-using-time?redirectedFrom=fulltext}
}

@incollection{casida1995time,
  title={Time-dependent density functional response theory for molecules},
  author={Casida, Mark E},
  booktitle={Recent Advances In Density Functional Methods: (Part I)},
  pages={155--192},
  year={1995},
  publisher={World Scientific},
  url = {https://www.worldscientific.com/doi/abs/10.1142/9789812830586_0005}
}

@article{adamo2013calculations,
  title={The calculations of excited-state properties with Time-Dependent Density Functional Theory},
  author={Adamo, Carlo and Jacquemin, Denis},
  journal={Chem. Soc. Rev.},
  volume={42},
  number={3},
  pages={845--856},
  year={2013},
  publisher={Royal Society of Chemistry},
  url = {https://pubs.rsc.org/en/content/articlelanding/2003/0n/c2cs35394f/unauth}
}

@article{hashmi2022valley,
  title={Valley polarization control in WSe 2 monolayer by a single-cycle laser pulse},
  author={Hashmi, Arqum and Yamada, Shunsuke and Yamada, Atsushi and Yabana, Kazuhiro and Otobe, Tomohito},
  journal={Phys. Rev. B},
  volume={105},
  number={11},
  pages={115403},
  year={2022},
  publisher={APS},
  url = {https://journals.aps.org/prb/abstract/10.1103/PhysRevB.105.115403}
}

@article{castro2004excited,
  title={Excited states dynamics in time-dependent density functional theory: high-field molecular dissociation and harmonic generation},
  author={Castro, Alberto and Marques, Miguel AL and Alonso, Julio A and Bertsch, George F and Rubio, Angel},
  journal={Eur. Phys. J. D},
  volume={28},
  pages={211--218},
  year={2004},
  publisher={Springer},
  url = {https://link.springer.com/article/10.1140/epjd/e2003-00306-3}
}

@article{yamada2019energy,
  title={Energy transfer from intense laser pulse to dielectrics in time-dependent density functional theory},
  author={Yamada, Atsushi and Yabana, Kazuhiro},
  journal={Eur. Phys. J. D},
  volume={73},
  pages={1--9},
  year={2019},
  publisher={Springer},
  url = {https://link.springer.com/article/10.1140/epjd/e2019-90334-7}
}

@article{tani2022enhanced,
  title={Enhanced energy deposition and carrier generation in silicon induced by two-color intense femtosecond laser pulses},
  author={Tani, Mizuki and Sasaki, Kakeru and Shinohara, Yasushi and Ishikawa, Kenichi L},
  journal={Phys. Rev. B},
  volume={106},
  number={19},
  pages={195141},
  year={2022},
  publisher={APS},
  url = {https://journals.aps.org/prb/abstract/10.1103/PhysRevB.106.195141}
}

@article{novoselov2004electric,
  title={Electric field effect in atomically thin carbon films},
  author={Novoselov, Kostya S and Geim, Andre K and Morozov, Sergei V and Jiang, De-eng and Zhang, Yanshui and Dubonos, Sergey V and Grigorieva, Irina V and Firsov, Alexandr A},
  journal={Science},
  volume={306},
  number={5696},
  pages={666--669},
  year={2004},
  publisher={American Association for the Advancement of Science},
  url = {https://www.science.org/doi/full/10.1126/science.1102896?casa_token=Od_1ZFviKlgAAAAA%3A02Yg0oiZnxRTyvIzTFmCnyiqN3KeDpwHXsxtre2_SIvweIt-kdTe0beLL_YzOh8AoFJFVlddk70a3IQ}
}

@article{geim2009graphene,
  title={Graphene: status and prospects},
  author={Geim, Andre Konstantin},
  journal={Science},
  volume={324},
  number={5934},
  pages={1530--1534},
  year={2009},
  publisher={American Association for the Advancement of Science},
  url = {https://www.science.org/doi/full/10.1126/science.1158877?casa_token=QJfGj4ya2pIAAAAA%3A-tPYn2itrgxXMwYbKFhfjYPe_wIOHMt1NCkKt4rZfkb80CV6126vjSjPMV9UWqi2Hh8vuIOXogoH9sQ}
}

@article{castro2009electronic,
  title={The electronic properties of graphene},
  author={Castro Neto, Antonio H and Guinea, Francisco and Peres, Nuno MR and Novoselov, Kostya S and Geim, Andre K},
  journal={Rev. of Mod. Phys.},
  volume={81},
  number={1},
  pages={109--162},
  year={2009},
  publisher={APS},
  url = {https://journals.aps.org/rmp/abstract/10.1103/RevModPhys.81.109}
}

@article{ishikawa2010nonlinear,
  title={Nonlinear optical response of graphene in time domain},
  author={Ishikawa, Kenichi L},
  journal={Phys. Rev. B},
  volume={82},
  number={20},
  pages={201402},
  year={2010},
  publisher={APS},
  url = {https://journals.aps.org/prb/abstract/10.1103/PhysRevB.82.201402}
}

@article{ishikawa2013electronic,
  title={Electronic response of graphene to an ultrashort intense terahertz radiation pulse},
  author={Ishikawa, Kenichi L},
  journal={New J. Phys.},
  volume={15},
  number={5},
  pages={055021},
  year={2013},
  publisher={IOP Publishing},
  url = {https://iopscience.iop.org/article/10.1088/1367-2630/15/5/055021/meta}
}

@article{fuchs1999ab,
  title={Ab initio pseudopotentials for electronic structure calculations of poly-atomic systems using density-functional theory},
  author={Fuchs, Martin and Scheffler, Matthias},
  journal={Comput. Phys. Commun.},
  volume={119},
  number={1},
  pages={67--98},
  year={1999},
  publisher={Elsevier},
  url = {https://www.sciencedirect.com/science/article/abs/pii/S001046559800201X}
}

@article{perdew1981self,
  title={Self-interaction correction to density-functional approximations for many-electron systems},
  author={Perdew, John P and Zunger, Alex},
  journal={Phys. Rev. B},
  volume={23},
  number={10},
  pages={5048},
  year={1981},
  publisher={APS},
  url = {https://journals.aps.org/prb/abstract/10.1103/PhysRevB.23.5048}
}

@article{runge1984density,
  title={Density-functional theory for time-dependent systems},
  author={Runge, Erich and Gross, Eberhard KU},
  journal={Phys. Rev. Lett.},
  volume={52},
  number={12},
  pages={997},
  year={1984},
  publisher={APS},
  url = {https://journals.aps.org/prl/abstract/10.1103/PhysRevLett.52.997}
}

@article{morimoto2024field,
  title={Field-Induced Rocking-Curve Effects in Attosecond Electron Diffraction},
  author={Morimoto, Yuya and Baum, Peter},
  journal={Phys. Rev. Lett.},
  volume={132},
  number={21},
  pages={216902},
  year={2024},
  publisher={APS},
  url = {https://journals.aps.org/prl/abstract/10.1103/PhysRevLett.132.216902} 
}

@article{morimoto2018diffraction,
  title={Diffraction and microscopy with attosecond electron pulse trains},
  author={Morimoto, Yuya and Baum, Peter},
  journal={Nat. Phys.},
  volume={14},
  number={3},
  pages={252--256},
  year={2018},
  publisher={Nature Publishing Group UK London},
  url = {https://www.nature.com/articles/s41567-017-0007-6}
}

@article{franco2007robust,
  title={Robust ultrafast currents in molecular wires through stark shifts},
  author={Franco, Ignacio and Shapiro, Moshe and Brumer, Paul},
  journal={Phys. Rev. Lett.},
  volume={99},
  number={12},
  pages={126802},
  year={2007},
  publisher={APS},
  url = {https://journals.aps.org/prl/abstract/10.1103/PhysRevLett.99.126802}
}

@article{schiffrin2013optical,
  title={Optical-field-induced current in dielectrics},
  author={Schiffrin, Agustin and Paasch-Colberg, Tim and Karpowicz, Nicholas and Apalkov, Vadym and Gerster, Daniel and M{\"u}hlbrandt, Sascha and Korbman, Michael and Reichert, Joachim and Schultze, Martin and Holzner, Simon and others},
  journal={Nature},
  volume={493},
  number={7430},
  pages={70--74},
  year={2013},
  publisher={Nature Publishing Group UK London},
  url = {https://www.nature.com/articles/nature11567}
}

@article{chen2018stark,
  title={Stark control of electrons along nanojunctions},
  author={Chen, Liping and Zhang, Yu and Chen, GuanHua and Franco, Ignacio},
  journal={Nat. Commun.},
  volume={9},
  number={1},
  pages={1--12},
  year={2018},
  publisher={Nature Publishing Group},
  url = {https://www.nature.com/articles/s41467-018-04393-4}
}

@article{heide2024petahertz,
  title={Petahertz electronics},
  author={Heide, Christian and Keathley, Phillip D and Kling, Matthias F},
  journal={Nat. Rev. Phys.},
  pages={1--15},
  year={2024},
  publisher={Nature Publishing Group UK London},
  url = {https://www.nature.com/articles/s42254-024-00764-7}
}

@article{yuan2024attosecond,
  title={Attosecond Diffraction Imaging of Electron Dynamics in Solids},
  author={Yuan, Mingrui and Golubev, Nikolay V},
  journal={arXiv preprint arXiv:2407.03537},
  year={2024},
  url = {https://arxiv.org/abs/2407.03537}
}

@article{heide2018coherent,
  title={Coherent electron trajectory control in graphene},
  author={Heide, Christian and Higuchi, Takuya and Weber, Heiko B and Hommelhoff, Peter},
  journal={Phys. Rev. Lett.},
  volume={121},
  number={20},
  pages={207401},
  year={2018},
  publisher={APS},
  url = {https://journals.aps.org/prl/abstract/10.1103/PhysRevLett.121.207401}
}

@article{falkovsky2007space,
  title={Space-time dispersion of graphene conductivity},
  author={Falkovsky, LA and Varlamov, AA},
  journal={Eur. Phys. J. B},
  volume={56},
  pages={281--284},
  year={2007},
  publisher={Springer},
  url = {https://link.springer.com/article/10.1140/epjb%2Fe2007-00142-3}
}

@article{kuzmenko2008universal,
  title={Universal optical conductance of graphite},
  author={Kuzmenko, Alexey B and Van Heumen, Erik and Carbone, Fabrizio and Van Der Marel, Dirk},
  journal={Phys. Rev. Lett.},
  volume={100},
  number={11},
  pages={117401},
  year={2008},
  publisher={APS},
  url = {https://journals.aps.org/prl/abstract/10.1103/PhysRevLett.100.117401}
}

@article{suzuki2017electronic,
  title={Electronic band structure of graphene based on the rectangular 4-atom unit cell},
  author={Suzuki, Akira and Tanabe, Masashi and Fujita, Shigeji},
  journal={J. Mod. Phys.},
  volume={8},
  number={4},
  pages={607--621},
  year={2017},
  publisher={Scientific Research Publishing},
  url = {https://www.scirp.org/journal/paperinformation?paperid=74995}
}

@article{taucer2017nonperturbative,
  title={Nonperturbative harmonic generation in graphene from intense midinfrared pulsed light},
  author={Taucer, Marco and Hammond, TJ and Corkum, PB and Vampa, Giulio and Couture, C and Thire, Nicolas and Schmidt, BE and L{\'e}gar{\'e}, F and Selvi, Hakan and Unsuree, Nawapong and others},
  journal={Phys. Rev. B},
  volume={96},
  number={19},
  pages={195420},
  year={2017},
  publisher={APS},
  url ={https://journals.aps.org/prb/abstract/10.1103/PhysRevB.96.195420}
}

@article{heide2021optical,
  title={Optical current generation in graphene: CEP control vs. $\omega$+ 2$\omega$ control},
  author={Heide, Christian and Boolakee, Tobias and Eckstein, Timo and Hommelhoff, Peter},
  journal={Nanophotonics},
  volume={10},
  number={14},
  pages={3701--3707},
  year={2021},
  publisher={De Gruyter},
  url = {https://www.degruyterbrill.com/document/doi/10.1515/nanoph-2021-0236/html}
}

@article{yabana1996time,
  title={Time-dependent local-density approximation in real time},
  author={Yabana, Kazuhiro and Bertsch, GF},
  journal={Phys. Rev. B},
  volume={54},
  number={7},
  pages={4484},
  year={1996},
  publisher={APS},
  url = {https://journals.aps.org/prb/abstract/10.1103/PhysRevB.54.4484}
}

@article{yabana2012time,
  title={Time-dependent density functional theory for strong electromagnetic fields in crystalline solids},
  author={Yabana, Kazuhiro and Sugiyama, T and Shinohara, Y and Otobe, T and Bertsch, GF},
  journal={Phys. Rev. B},
  volume={85},
  number={4},
  pages={045134},
  year={2012},
  publisher={APS},
  url = {https://journals.aps.org/prb/abstract/10.1103/PhysRevB.85.045134}
}

@article{kaneshima2018polarization,
  title={Polarization-resolved study of high harmonics from bulk semiconductors},
  author={Kaneshima, Keisuke and Shinohara, Yasushi and Takeuchi, Kengo and Ishii, Nobuhisa and Imasaka, Kotaro and Kaji, Tomohiro and Ashihara, Satoshi and Ishikawa, Kenichi L and Itatani, Jiro},
  journal={Phys. Rev. Lett.},
  volume={120},
  number={24},
  pages={243903},
  year={2018},
  publisher={APS},
  url = {https://journals.aps.org/prl/abstract/10.1103/PhysRevLett.120.243903}
}

@article{weitz2024strong,
  title={Strong-field Bloch electron interferometry for band-structure retrieval},
  author={Weitz, Tobias and Heide, Christian and Hommelhoff, Peter},
  journal={Phys. Rev. Lett.},
  volume={132},
  number={20},
  pages={206901},
  year={2024},
  publisher={APS},
  url ={https://journals.aps.org/prl/abstract/10.1103/PhysRevLett.132.206901}
}

@article{kelardeh2015graphene,
  title={Graphene in ultrafast and superstrong laser fields},
  author={Kelardeh, Hamed Koochaki and Apalkov, Vadym and Stockman, Mark I},
  journal={Phys. Rev. B},
  volume={91},
  number={4},
  pages={045439},
  year={2015},
  publisher={APS},
  url = {https://journals.aps.org/prb/abstract/10.1103/PhysRevB.91.045439}
}

@article{hashmi2023enhancement,
  title={Enhancement of valley-selective excitation by a linearly polarized two-color laser pulse},
  author={Hashmi, Arqum and Yamada, Shunsuke and Yabana, Kazuhiro and Otobe, Tomohito},
  journal={Phys. Rev. B},
  volume={107},
  number={23},
  pages={235403},
  year={2023},
  publisher={APS},
  url = {https://journals.aps.org/prb/abstract/10.1103/PhysRevB.107.235403}
}

@article{hashmi2022nonlinear,
  title={Nonlinear dynamics of electromagnetic field and valley polarization in WSe2 monolayer},
  author={Hashmi, Arqum and Yamada, Shunsuke and Yamada, Atsushi and Yabana, Kazuhiro and Otobe, Tomohito},
  journal={Appl. Phys. Lett.},
  volume={120},
  number={5},
  year={2022},
  publisher={AIP Publishing},
 url = {https://pubs.aip.org/aip/apl/article/120/5/051108/2832785}
}

@article{otobe2024influence,
  title={Influence of point defects on laser-induced excitation in silicon},
  author={Otobe, Tomohito and Gushiken, Eiyu},
  journal={Phys. Rev. Applied},
  volume={22},
  number={6},
  pages={064096},
  year={2024},
  publisher={APS},
  url ={https://journals.aps.org/prapplied/abstract/10.1103/PhysRevApplied.22.064096}
}

@article{tani2021semiclassical,
  title={Semiclassical description of electron dynamics in extended systems under intense laser fields},
  author={Tani, Mizuki and Otobe, Tomohito and Shinohara, Yasushi and Ishikawa, Kenichi L},
  journal={Phys. Rev. B},
  volume={104},
  number={7},
  pages={075157},
  year={2021},
  publisher={APS},
  url = {https://journals.aps.org/prb/abstract/10.1103/PhysRevB.104.075157}
}

@article{bertsch2000real,
  title={Real-space, real-time method for the dielectric function},
  author={Bertsch, George F and Iwata, J-I and Rubio, Angel and Yabana, Kazuhiro},
  journal={Physical Review B},
  volume={62},
  number={12},
  pages={7998},
  year={2000},
  publisher={APS},
  url = {https://journals.aps.org/prb/abstract/10.1103/PhysRevB.62.7998}
}

@article{kleinman1982efficacious,
  title={Efficacious form for model pseudopotentials},
  author={Kleinman, Leonard and Bylander, DM},
  journal={Physical review letters},
  volume={48},
  number={20},
  pages={1425},
  year={1982},
  publisher={APS},
  url = {https://journals.aps.org/prl/abstract/10.1103/PhysRevLett.48.1425}
}

@article{yamada2025saturable,
  title={Saturable absorption in highly excited laser-irradiated silicon and its suppression at the surface},
  author={Yamada, Shunsuke and Otobe, Tomohito},
  journal={Phys. Rev. B},
  volume={111},
  number={7},
  pages={075105},
  year={2025},
  publisher={APS},
  url = {https://journals.aps.org/prb/abstract/10.1103/PhysRevB.111.075105}
}

@article{xu2011giant,
  title={Giant surface charge density of graphene resolved from scanning tunneling microscopy and first-principles theory},
  author={Xu, P and Yang, Y and Barber, SD and Ackerman, ML and Schoelz, JK and Kornev, Igor A and Barraza-Lopez, Salvador and Bellaiche, L and Thibado, PM},
  journal={Phys. Rev. B},
  volume={84},
  number={16},
  pages={161409},
  year={2011},
  publisher={APS},
  url = 
{https://journals.aps.org/prb/abstract/10.1103/PhysRevB.84.161409}
}

@article{houston1940acceleration,
  title={Acceleration of electrons in a crystal lattice},
  author={Houston, WV},
  journal={Phys. Rev.},
  volume={57},
  number={3},
  pages={184},
  year={1940},
  publisher={APS},
  url = 
{https://journals.aps.org/pr/abstract/10.1103/PhysRev.57.184}
}

@PREAMBLE{
 "\providecommand{\noopsort}[1]{}" 
 # "\providecommand{\singleletter}[1]{#1}%" 
}

\end{document}